\documentclass[12pt]{iopart}

\usepackage{graphicx}
\begin{document}

\title {Realistic shell-model calculations: current status and open problems}

\author{A Covello$^{1,2}$ and A Gargano$^2$}

\address{$^1$Dipartimento di Scienze Fisiche, Universit\`{a} di Napoli Federico II, \\
Complesso Universitario di Monte S. Angelo, I-80126 Napoli, Italy \\
$^2$Istituto Nazionale di Fisica Nucleare, \\
Complesso Universitario di Monte S. Angelo, I-80126 Napoli, Italy\\}

\ead{covello@na.infn.it}

\begin{abstract}
The main steps involved in realistic shell-model calculations employing two-body low-momentum interactions are briefly reviewed. The practical value of this approach is exemplified by the results of recent calculations and some remaining open questions and directions for future research are discussed.

\end{abstract}

\maketitle

\section{Introduction}

A key challenge for nuclear physics is to develop a comprehensive understanding of the observed properties of nuclei starting from the forces among nucleons. Nowadays there are two main ways to try to achieve this goal, which are substantially complementary to each other.  
The first one is based on the so-called {\it ab initio} calculations in which nuclear properties, such as binding and excitation energies, are calculated directly from first principles of quantum mechanics using an appropriate computational scheme. To this category belong the Green's function Montecarlo Method (GFMC) \cite{Pieper08}, the no-core shell-model (NCSM) \cite{Navratil09}, and the coupled cluster method \cite{Hagen07}. 
Clearly, all {\it ab initio} calculations need huge amount of computational resources and are therefore currently limited to light nuclei. 

The second line of attack consists of using the shell model with realistic two-body effective interactions, namely derived microscopically from  the free nucleon-nucleon ($NN$) potential. In this kind of calculations only the valence nucleons are treated as active particles while the core polarization effects are taken  into account  perturbatively in the derivation of the effective interaction $V_{\rm eff}$. Of course, this approach allows one to perform calculations for  medium- and heavy-mass nuclei which are far beyond the reach of {\it ab initio} calculations.

In this context, it should be mentioned that in recent years successful shell-model calculations, in particular for $sd$- and $pf$-nuclei, have been performed \cite{Brown01, Caurier05,Otsuka01} by employing modified versions of realistic effective interactions. Let us make it clear that this paper is concerned only with calculations employing  two-body effective interactions derived from the  free $NN$ potential without any phenomenological adjustments, which we refer to as realistic shell-model calculations {\it tout court}.

The two main ingredients of realistic shell-model calculations are the $NN$ potential 
$V_{NN}$ and the many-body methods for deriving $V_{\rm eff}$. As regards the latter, a main difficulty with its derivation is the strong short-range repulsion contained in modern $NN$ potentials. As is well known, the traditional way to circumvent this problem is the Brueckner $G$-matrix method. However, a few years ago a new approach to the renormalization of $V_{NN}$  has been proposed \cite{Bogner02,Bogner03}, which has proved to be \cite{Coraggio09a} an advantageous alternative to the use of the $G$-matrix . This consists in constructing a smooth low-momentum potential, $V_{\rm low-k}$, that can be used directly to derive  $V_{\rm eff}$ within the framework of a perturbative approach. 

Since its introduction in 2002, we have used this approach to derive realistic low-momentum effective interactions for shell-model calculations in various mass regions, which have  provided an accurate description of nuclear structure properties. It should be noted that, having taken the single-nucleon energies from experiment, these calculations are essentially parameter-free. 

While the realistic shell-model calculations performed thus far have all been based on a Hamiltonian containing only two-body interactions, the role of three-nucleon interactions
has been deeply investigated within the framework of {\it ab initio} approaches, such as  
the GFMC and the NCSM. It is certainly a major open problem to investigate the contribution of three-body forces to the shell-model effective interaction.

In this paper, we shall try to point out some questions that in our opinion remain open in  
shell-model calculations with realistic effective interactions. This is done in the context of a brief description of the current status of the field, which serves the purpose to provide the appropriate frame for our discussion.

\section{Theory and applications}

The Schr\"odinger equation for a nucleus with A nucleons is written as

\begin{equation}
H\Psi_i=(H_{0}+H_{1})\Psi_i = E_i \Psi_i, \label{Schr} \end{equation}

\noindent
 where
\begin{equation}
H_{0} = T+U
\label{defh0}
\end{equation}
\noindent
and
\begin{equation}
H_{1}= V_{NN}-U,
\label{defh1}
\end{equation}

\noindent
$T$ being the kinetic energy and $U$ an auxiliary  potential introduced to define a convenient single-particle basis.
The effective interaction $V_{\rm eff}$ acting only within a reduced model space is then defined through the eigenvalue problem

\begin{equation}
PH_{\rm eff}P| \Psi_\alpha\rangle = P(H_{0}+V_{\rm eff})P| \Psi_\alpha\rangle= E_\alpha P \Psi_\alpha, \label{defheff} \end{equation}

\noindent
where the  $E_\alpha$  and the corresponding $\Psi_\alpha$ are a subset of the
eigenvalues  and eigenfunctions of the original Hamiltonian defined in the complete Hilbert space. The $P$ operator projects onto the chosen model space, which is defined in terms of the eigenvectors of the unperturbed Hamiltonian $H_{0}$.

A well-established approach to the derivation of realistic effective interactions is provided by the $\hat Q$-box folded-diagram expansion \cite{Coraggio09a}.
It is, however, outside the scope of this article to give any detailed discussion of this approach. Rather, our aim is to discuss synthetically its merits and limits. To this end, this section is divided into two parts. In the first one we examine in a critical way the main steps involved in the derivation of $V_{\rm eff}$. In the second one we give a brief survey of some applications to show the accuracy of realistic shell-model calculations in the description of nuclear structure properties.

\subsection{Outline of theoretical framework}

\subsubsection{Choice of $V_{NN}$}

Starting in the mid 1990s, several potential models have been constructed which predict almost identical deuteron observables and are phase-shift equivalent, namely they fit equally well ($\chi^2/{\rm datum} \approx 1$) the $NN$ scattering data up to the inelastic threshold. These are the potentials constructed by the Nijmegen group, Nijm I and Nijm II 
\cite{Stocks94}, the Argonne $V_{18}$ potential \cite{Wiringa95}, and the CD-Bonn potential \cite{Machleidt01}. 

More recently, new $NN$ potentials have been derived within the framework of the chiral perturbation theory following the basic idea of Weinberg \cite{Weinberg90}. Efforts in this direction have resulted in the construction of next-to-next-to-next-to-leading order (N$^3$LO, fourth order) potentials \cite{Entem03,Epelbaum05}. An interesting feature of this approach is that three-nucleon forces  arise naturally at third order (N$^2$LO). 

We will not go further into this subject and refer the reader to \cite{Machleidt07} for a survey of the current status of the chiral potentials. However, a general comment may be in order here. Currently there are no real efforts to go beyond the one-boson-exchange (OBE) model in the derivation of $V_{NN}$ within the traditional meson theory, the main interest in the field being focused on chiral potentials. As regards the latter, there remain some open problems,  the most important ones being the three-nucleon forces  beyond N$^2$LO \cite{Machleidt09} and the consistency of the Weinberg power counting \cite{Entem08}.

Of course, the existence of several different phase-shift equivalent $NN$ potentials raises the question of how much nuclear structure results may depend on the  choice of the potential one starts with. 
However, the renormalization of the free $NN$ potential, which is the subject of section 2.1.2, removes to a large extent the ambiguity in the choice of $V_{NN}$. We have indeed verified \cite{Coraggio09a} that low-momentum shell-model effective interactions derived from phase-shift equivalent $NN$ potentials through the $V_{\rm low-k}$ approach do not lead to significantly different results.

\subsubsection{Renormalization of $V_{NN}$}

In recent years, a renormalization-group-based approach has been introduced 
\cite{Bogner02,Bogner03} allowing one to handle the nonperturbative behaviour that characterizes all realistic $NN$ potential models. In this approach, the  high-momentum modes of $V_{NN}$ are integrated out down to a certain cutoff momentum $\Lambda$,  which  leads to a low-momentum potential that decouples high- and low-energy degrees of freedom while preserving  the on-shell properties of the original $V_{NN}$. By varying the cutoff, an entire class of low-momentum potentials can be associated with the initial Hamiltonian, which are all by construction softer than  $V_{NN}$. This greatly simplifies, or even makes feasible in some cases, nuclear structure calculations.
It is a remarkable feature of this approach that with a cutoff in the range $\Lambda\sim 2.1 \;{\rm fm}^{-1}$ low-momentum interactions derived from different $NN$ potential become very similar to each other. This has recently led to a widespread use of these low-momentum potentials in various contexts, going from few-body systems to nuclear matter.

Different techniques have been developed to derive low-momentum potentials, the decimation of the initial  potential being performed on the two-body systems in all cases.
In our calculations, we have adopted a conventional effective interaction technique based on the Lee-Suzuki similarity transformation \cite{Suzuki80}. The cutoff $\Lambda$ specifies the  low-momentum space $\mathcal{P}$ within which the $V_{\rm low-k}$  is confined and its complement $\mathcal{Q}=1- \mathcal{P}$.
Then, using  a similarity transformation $X$ a new  Hamiltonian $\mathcal{H}= X^{-1} H X$ is defined which satisfies the decoupling equation

\begin{equation}
\mathcal{Q} \mathcal{H} \mathcal{P}=0.
\label{dec}
\end{equation}

\noindent
The low-momentum potential is then given by

\begin{equation}
V_{\rm low-k} = \mathcal{P} \mathcal{H} \mathcal{P} - \mathcal{P}T \mathcal{P}.
\label{vlow}
\end{equation}

\noindent The solution procedure for Eq. (5) can be found in \cite{Coraggio09a}. Here we only mention that we adopt the iterative technique for non-degenerate model spaces proposed  in \cite{Andreozzi96} and  employ a momentum-space discretization procedure with an adequate number of Gaussian mesh points.

As mentioned above, the $V_{\rm low-k}$ is developed for the two-body system, which means that when used in calculations  for $A>2$ the low-energy observables are not the same as those produced by the original $NN$ potential and  depend, to a certain extent, on the value of the cutoff \cite{Nogga04}. This dependence may be removed by complementing, for each value of $\Lambda$, the two-body  $V_{\rm low-k}$ potential with three- and higher-body components, which are in fact generated by the renormalization procedure when applied to an $A$-body system, even if the initial Hamiltonian contains only a two-body potential \cite{Bogner09}. However, the three-body forces included in $V_{\rm low-k}$ calculations to date are generally approximated by fitting the parameters entering the leading chiral three-nucleon forces. A consistent evolution of the three-body forces in the three-nucleon system has been only recently achieved in the work of \cite{Jurgenson09}. There, calculations are performed for $A\leq4$ nuclei with an
initial Hamiltonian including as two- and three-body forces the
N$^{3}$LO \cite{Entem03} and N$^{2}$LO  \cite{Gazit09} interactions, respectively. 

Note that the inclusion of forces beyond the two-body ones has been essentially limited to few-nucleon systems. As regards medium- and heavy-mass nuclei,  only the $NN$ potential is used at present in the derivation of the shell-model effective interaction. We shall come back to this point in section 3.

It is worth mentioning that  in almost all our calculations the cutoff parameter $\Lambda$ has been chosen in the vicinity of 2 fm$^{-1}$. This means that $V_{\rm low-k}$ preserves the phase shifts up to $E_{\rm Lab} \approx 350$ MeV, which is the inelastic threshold. We have found  \cite{Covello05}, however, that the shell-model results do not  change significantly for moderate variations of the cutoff around this value. We have also verified \cite{Covello05} that allowing for limited changes ($\sim 0.3 \; {\rm fm}^{-1}$) in the value of $\Lambda$ the $V_{\rm low-k}$'s extracted from different phase-shift equivalent potentials give essentially the same results. Further study is certainly needed to better clarify the effect of cutoff variations on shell-model effective interactions.

In concluding this section, it is worth  insisting on the fact that the use of $V_{\rm low-k}$ rather than the  $G$ matrix represents a substantial progress in the derivation of shell-model effective interactions. The main advantages of $V_{\rm low-k}$ may be summarized as follows:
i) since it is a smooth $NN$ potential, $V_{\rm low-k}$ can be used directly in nuclear structure calculations within a perturbative approach; ii) it does not depend either on the starting energy or on the model space, as instead the case of the $G$ matrix, which is defined in the nuclear medium; iii) the $V_{\rm low-k}$'s extracted from various phase-shift equivalent potentials are very similar to each other, thus suggesting the realization of a nearly unique low-momentum potential.

From the above, it is clear that $V_{\rm low-k}$ is also a most valuable tool to tackle the tough problem of taking into account three-body forces in realistic shell-model calculations.

\subsubsection{Derivation of the effective interaction}

As mentioned above, the calculation of the effective interaction within a chosen model space $P$ is performed by using the $\hat{Q}$-box folded-diagram expansion developed for two valence particles.

The effective interaction of Eq.~(\ref{defheff}) can be written as

\begin{equation}
V_{\rm eff} = \hat{Q} + \sum_{i=1}^{\infty} F_{i}~, \label{fold} \end{equation}

\noindent
where the $\hat{Q}$-box is defined  as the sum of all irreducible and valence linked diagrams with  $V_{\rm low-k}$ replacing $V_{NN}$ in the $H_{1}$ vertices, and the $F_i$'s  represent $\hat{Q}$-box $i$-folded diagrams. The latter can be  expressed in terms of $\hat{Q}$-box derivatives with respect to the energy variable $\omega$, namely

\begin{eqnarray}
F_1 &=& \hat{Q}_1 \hat{Q},\nonumber\\
F_2 &=& \hat{Q}_2 \hat{Q} \hat{Q} + \hat{Q}_1 \hat{Q}_1 \hat{Q},\nonumber\\
F_3 &=& \hat{Q}_3  \hat{Q} \hat{Q} \hat{Q} + \hat{Q}_2 \hat{Q}_1 \hat{Q} \hat{Q} +
\hat{Q}_2 \hat{Q} \hat{Q}_{1} \hat{Q} +\hat{Q}_1 \hat{Q}_2 \hat{Q} \hat{Q}+
\hat{Q}_1 \hat{Q}_{1} \hat{Q}_{1} \hat{Q}, \label{fm} \\ 
... ~,\nonumber \end{eqnarray}

\noindent
with

\begin{equation}
\hat{Q}_m = \frac {1}{m!} \frac {d^m \hat{Q} (\omega)}{d \omega^m} \biggl| _{\omega=\omega_0},\label{qm} \end{equation}

\noindent
$\omega_0$ being the unperturbed energy of two valence particles in a degenerate model space, $PH_{0}P=\omega_{0}$.

As a first step, one has to calculate the $\hat{Q}$-box. This calculation requires some approximations, since only  diagrams up to a finite order in $H_1$ can be included and a truncation of the intermediate-state summation has to be made for each $\hat{Q}$-box diagram. Then, for a given $\hat{Q}$-box the folded-diagram series~(\ref{fold}) can be summed
up by the  Lee-Suzuki iteration method \cite{Suzuki80}.

One of the most important questions arising in this kind of approach  is the convergence of the effective interaction expansion, to which much attention has been paid over the years.  Before folded diagrams were taken into account, both the order-by-order~\cite{Barrett70} and the intermediate state~\cite{Sommermann81} convergence were questioned.
Furthermore, it was shown~\cite{Schucan72} that the presence of an intruder state in the low-lying spectrum of the nucleus with two valence nucleons is an essential source of divergence. The problem of the intermediate state convergence was substantially solved by using  $NN$ potentials with a weak tensor force, while  the introduction of folded diagrams and their summation through the Lee-Suzuki scheme was proved to be an important step to handle the  non-convergence of the order-by-order expansion of $V_{\rm eff}$, also when intruder states are present in the $P$-space~\cite{Suzuki80,Shurpin83}.
As a matter of fact, if we know the $\hat{Q}$-box and its derivatives, $V_{\rm eff}$ is obtained after a small number of iterations. The convergence of the $\hat{Q}$-box, however, still remains to some extent an open problem, although it has been shown that  the differences in 
$V_{\rm eff}$'s calculated with $\hat{Q}$-boxes to various orders, are attenuated  when the folded diagrams are summed to all orders~\cite{Hjorth95}. In particular, in \cite{Hjorth95}
a comparative study between results obtained using a  second- and third-order $\hat{Q}$-box in the folded-diagram expansion has been performed.
There it was  found that both calculations give very similar results for nuclei with two valence nucleons, which, as it was also suggested in~\cite{Shurpin83}, should be traced to the cancellation of the higher-order $\hat{Q}$-box diagrams by the folded diagrams. However, the final conclusion of~\cite{Hjorth95} was that third-order calculations should be employed since third-order terms of the $\hat{Q}$-box are not negligible. On the other hand, in  almost all our calculations we have used effective interactions derived by including diagrams up to second order. It is worth mentioning that our choice is also  supported by the results of 
~\cite{Holt05}. There, a $V_{\rm low-k}$ derived from the CD-Bonn $NN$ potential has been employed and the $sd$-shell effective interaction has been calculated by summing core-polarization diagrams to all orders. The results of this calculation have turned  out to be remarkably similar to those of a second-order one.

There are a few other points that we would like to discuss here.
The effective interaction constructed within the above approach contains both one- and two-body terms. All the  one-body contributions, once summed to the eigenvalues of $H_0$, give the single-particle  effective energies, which should represent the energies of the 
one-valence-nucleon system.  In most of the realistic shell-model calculations to date, these  single-particle energies are replaced by the experimental values, and only the two-body term of $V_{\rm eff}$ is retained  by employing the subtraction procedure described in \cite{Shurpin83}.

The effective interaction is constructed for a two-valence-particle nucleus but is generally used also for more complex systems. This means that the three- or higher-body forces, which appear in the effective interaction for the latter, even if the initial potential is only a two-body one, are not taken into account. In \cite{Zuker03} the lack of three-body forces, effective or original, was considered responsible for the failure   to explain  simultaneously binding energies and excitation spectra of several nuclei. In~\cite{Ellis05}, using the two-level Lipkin model, it was shown that effective three-body forces are not negligible. On the basis of our calculations, however, we do not have at the moment a definite conclusion  on the role of these forces.  In the next section we shall comment on this point.

Finally, it is worth  recalling that the perturbation in our starting Hamiltonian is  $H_{1}=V_{\rm low-k}-U$, and, as a consequence,  the $\hat{Q}$-box diagrams contain $-U$ as well as  $V_{\rm low-k}$ vertices. However, in almost all calculations performed so far diagrams with the $U$ potential and bubble insertions are not included. This assumption is based on the fact that these diagrams would cancel each other when using for  $U$ the Hartree-Fock self-consistent potential. The most common choice, however, is the harmonic oscillator potential, and the plausibility of the above approximation remains to be ascertained. In some of our recent papers~\cite{Coraggio09d,Coraggio09e} all the $V_{\rm low-k}$ and $-U$ diagrams up to third order are taken into account.

From this discussion it seems that some  questions inherent  in the derivation of the effective interaction still remain to be settled, especially as regards the real advantages of an {\it a priori} more complete and  consistent calculation of the effective interaction. We feel  that further investigations aimed at finding  specific answers to these  questions are needed.
However, these  considerations cannot ignore the remarkable success achieved thus far by realistic shell-model calculations in explaining nuclear structure properties.
This will be illustrated in section 2.2.

\subsection{Practical value of realistic shell-model calculations}

While the first efforts to derive the shell-model effective interaction from the free $NN$ potential go back to more than forty years ago, the practical value of this approach has emerged only during the last decade or so. We shall here try to give a concise yet quantitative evaluation of the results produced by modern realistic shell-model calculations. These results concern spectroscopic properties of nuclei in various mass regions both near and far from stability, and have been obtained using second-order folded-diagram effective interactions derived from the CD-Bonn potential renormalized by constructing a $V_{\rm low-k}$with $\Lambda = 2.1-2.2$ fm$^{-1}$. In particular, calculations have been performed in the regions around doubly magic $^{100}$Sn, $^{132}$Sn and $^{208}$Pb involving some twenty nuclei altogether (\cite{Coraggio09a} and references therein, \cite{Coraggio09b,Coraggio09c}).
In practically all cases a very good agreement between the calculated and experimental spectra has been obtained. In general, the rms deviation does not exceed 100 keV. Also, the currently available experimental data on electromagnetic transition rates are well reproduced by the theory (see, for instance, \cite{Covello07}). 

It is well outside the scope of this article to give even a brief survey of the results of the realistic shell-model calculations referred to above. However, to quantitatively illustrate their practical value, we feel it appropriate to just discuss in some detail the results for the two odd-odd nuclei $^{136}$Sb and $^{134}$I. In both cases the main motivation for our calculations \cite{Simpson07,Coraggio09c} was to interpret new data of primary interest for the understanding of the shell structure of  nuclei in the vicinity of $^{132}$Sn. Note that $^{136}$Sb with an $N/Z$ ratio of 1.67 is at present the most exotic open-shell nucleus beyond $^{132}$Sn for which information exists on excited states.

\begin{figure}[h]
\begin{center}
\includegraphics[width=8cm]{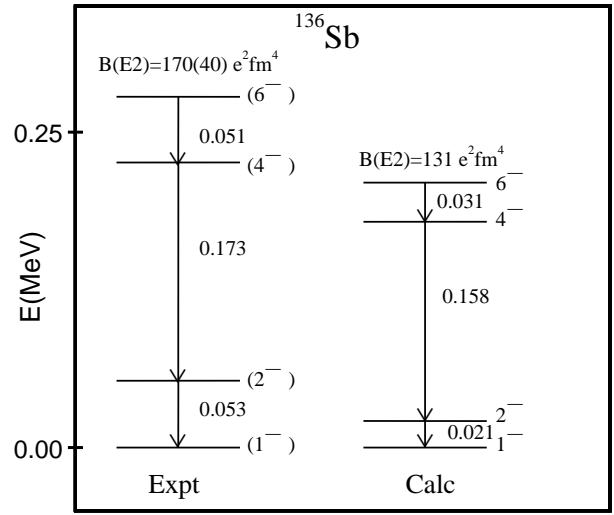}
\end{center}
\caption{\label{fig1} Experimental and calculated level scheme of $^{136}$Sb.}
\end{figure}

In the study \cite{Simpson07} two new transitions, in addition to the one previously known, were observed in $^{136}$Sb using $\gamma$-ray and conversion-electron spectroscopy, and a level scheme was constructed by comparison with the results of a realistic shell-model calculation performed within the theoretical framework described in the previous sections. As is seen in figure 1, the predicted energies are in very good agreement with measured values. The theory predicted also a $B(E2;6^- \rightarrow 4^-)$ value which comes quite close to the experimental one.

As regards the nucleus $^{134}$I,  five high-spin excited states up to an energy of about 3 MeV were very recently identified \cite{Liu09} through measurements of prompt $\gamma$ rays from the spontaneous fission of $^{252}$Cf. Angular correlations were performed, but were not sufficient to assign spins and parities. We have interpreted the observed states on the basis of a realistic shell-model calculation and made spin-parity assignments accordingly \cite{Coraggio09c}. As is seen in figure 2, a very good agreement was found between the experimental and calculated levels. 
\begin{figure}[h]
\begin{center}
\includegraphics[width=8cm]{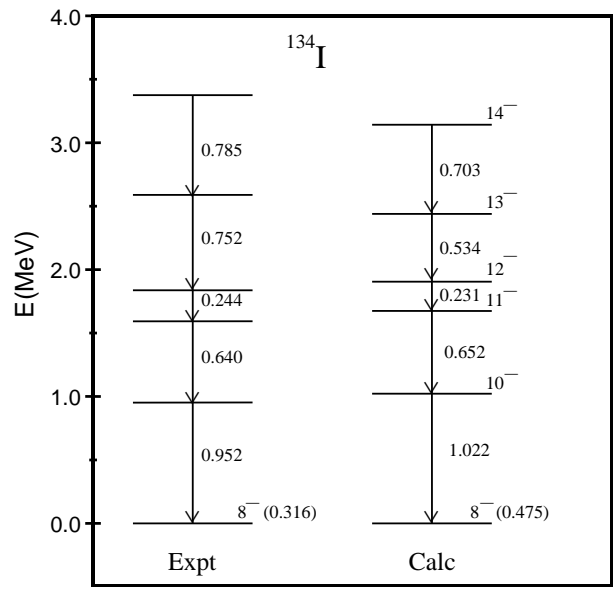}
\end{center}
\caption{\label{fig1} High-spin states in $^{134}$I identified in \cite{Liu09} compared with the calculated level scheme for negative-parity yrast states.}
\end{figure}

\section{More on open problems}

In the previous sections we have already touched upon some interesting questions regarding realistic shell-model calculations.  In this section, we give a more detailed discussion of 
what we feel are the issues worthy of further study in the near future.

The two examples given in section 2.2 provide good evidence of the predictive power of realistic shell-model calculations. However, as is the case for the majority of our calculations, they concern nuclei neighboring a double shell closure. It is, of course, a very interesting problem to find out what happens with increasing number of valence nucleons. From this viewpoint, of special interest are long isotopic or isotonic chains which provide the opportunity to investigate the effect of adding neutrons or protons to a doubly magic core over a large number of nuclei. To this aim, we have conducted a study of the $N=82$ isotonic chain which has long been considered a benchmark for shell-model calculations employing effective interactions derived from free $NN$ potentials \cite{Covello97,Holt97}. The results of this study \cite{Coraggio09a} have shown remarkable agreement with experiment for the low-energy spectra all over the isotonic chain. However, the ground-state energy, which is well reproduced for the two-valence-proton nucleus $^{134}$Te, is increasingly overestimated with increasing number of protons. It should mentioned that a similar behavior was found in \cite{Engeland98} for the Sn isotopic chain. This study, however, differs from that of \cite{Coraggio09a} in two respects:
the short-range repulsion of $V_{NN}$ is taken care of by introducing the Brueckner $G$ matrix and diagrams through third order in $G$ are included in the $\hat Q$-box. 
In \cite{Engeland98} it was shown that the calculated binding energies turned out to be in good agreement with experiment by adding a small repulsive monopole contribution to the effective two-body interaction. 

These findings lend support to the suggestion that the overbinding of nuclei with more than two valence particles produced by realistic shell-model calculations may be traced to the lack of many-body forces, in particular the three-body ones. However, we have recently performed calculations for oxygen \cite{Coraggio07} and calcium \cite{Coraggio09d} isotopes, as well as for  $N=82$ isotones \cite{Coraggio09e}, including diagrams up to third order in $V_{\rm low-k}$. The results obtained  seem to indicate that third-order diagrams may play a significant role in curing the above ``overbinding problem". 

In summary, we may say that a deeper comprehension of the role of both genuine and effective three-body forces and of contributions to the $\hat Q$-box beyond second order certainly deserves further study. In this connection, it should be mentioned that an attempt to explain the overbinding in neutron-rich O isotopes in terms of genuine three-body forces is made in \cite{Otsuka09}.

In the calculations for $^{136}$Sb and $^{134}$I mentioned above only particle-particle and particle-hole matrix elements come into play. We have, however, also performed several calculations for nuclei with two or more valence holes \cite{Genevey03,Scherillo04}. While the overall agreement between theory and experiment may be considered satisfactory in most cases, it is not in general of the same quality as that obtained for nuclei in which there is no hole-hole interaction. At present, we don't have a clear understanding of the origin of this dissimetry.

\section {Final remarks}

In this paper, we have tried to give an outline of the present status of shell-model calculations employing realistic two-body effective interactions  and point out some problems that deserve further study.
To place our discussion in a proper perspective, we have briefly surveyed the main steps
involved in this kind of calculations, whose practical value has been exemplified by the results of two recent calculations in the neutron-rich $^{132}$Sn region. 

As regards open problems, we have focused on those issues that are raised by our own studies. This may have left out other interesting questions. This is certainly the case 
of the coupling with the scattering continuum which should play an important role in nuclei
near the drip lines. We refer to \cite{Tsukiyama09} for a very recent study of this problem including references to earlier works. We hope, however, to have succeeded
in highlighting the value of realistic shell-model calculations and the need for further efforts to clarify some remaining open questions.

\section*{Acknowlegments}
The considerations made in this paper are essentialy the fruit of a long-standing collaboration with Luigi Coraggio, Nunzio Itaco and Tom Kuo. Frequent discussions with Luigi Coraggio and Nunzio Itaco have been of great help.


\section*{References}
\begin{thebibliography}{10}
\bibitem{Pieper08} Pieper S 2008 {\it Proc. Int. School of Physics
``E. Fermi", Course CLXIX}, ed A Covello, F Iachello, R A Ricci and G Maino (Amsterdam: IOS Press) p 111 and references therein
\bibitem{Navratil09}  Navr\'atil P, Quaglioni S, Stetcu I and Barrett B R 2009 {\it J. Phys. G: Nuc. Part. Phys.} {\bf 36} 083101 and references therein
\bibitem{Hagen07}  Hagen G, Dean D J, Hjorth-Jensen M, Papenbrock T and Schwenk A 2007 {\it Phys. Rev.} C {\bf 76} 044305 and references therein
\bibitem{Brown01} Brown B A 2001 {\it Prog. Part. Nucl. Phys.} {\bf 47} 517 
\bibitem{Caurier05} Caurier E, Martinez-Pinedo G, Nowacki F, Poves A and Zuker A P 2005
{\it Rev. Mod. Phys.} {\bf 77} 427 
\bibitem{Otsuka01} Otsuka T, Honma M, Mizusaki T and Shimizu N  2001 {\it Prog. Part. Nucl. Phys.} {\bf 47} 319
\bibitem{Bogner02} Bogner S, Kuo T T S, Coraggio L, Covello A and Itaco N 2002 {\it Phys. Rev.} C {\bf 65} 051301(R)
\bibitem{Bogner03} Bogner S, Kuo T T S and Schwenk A 2003 {\it Phys. Rep. } {\bf 386} 1
\bibitem{Coraggio09a} Coraggio L, Covello A, Gargano A, Itaco N and Kuo T T S 2009 {\it Prog. Part. Nucl. Phys.} {\bf 62} 135
\bibitem{Stocks94} Stocks V G J, Klomp A M, Terheggen C P F and de Swart J 1994 
{\it Phys. Rev. }  C {\bf 49} 2950
\bibitem{Wiringa95} Wiringa R B, Stoks V G J and Schiavilla R 1995 
{\it Phys. Rev. } C {\bf 51} 38
\bibitem{Machleidt01} Machleidt R 2001 {\it Phys. Rev.} C {\bf 63} 024001
\bibitem{Weinberg90} Weinberg S 1990 {\it Phys. Lett.} B {\bf 251} 288; 1991 
{\it Nucl. Phys.} B {\bf 363} 3
\bibitem{Entem03} Entem D R and  Machleidt R 2003 {\it Phys. Rev.} C {\bf 68} 041001(R)
\bibitem{Epelbaum05} Epelbaum E, Gl\"ockle W and Meissner  U-G 2005 {\it Nucl. Phys.} A {\bf 747} 362
\bibitem{Machleidt07} Machleidt R 2007 {\it Nucl. Phys.} A  {\bf 790} 17c
\bibitem{Machleidt09} Machleidt R 2009 arXiv:nucl-th/0909.2881
\bibitem{Entem08} Entem D R, Ruiz Arriola E, Pav\'on Valderrama M and Machleidt R 2008 {\it Phys. Rev.} C {\bf 77} 044006 and references therein
\bibitem{Suzuki80} Suzuki K and Lee S Y 1980  {\it Prog. Theor. Phys.} {\bf 64} 2091
\bibitem{Andreozzi96} Andreozzi F 1996 {\it Phys. Rev. } C {\bf 54} 684
\bibitem{Nogga04} Nogga A, Bogner S K and Schwenk A 2004 {\it Phys. Rev.} C {\bf 70} 061002(R)
\bibitem{Bogner09} Bogner S K, Furnstahl R J and Schwenk A 2009 arXiv:nucl-th/0912.3688
\bibitem{Jurgenson09} Jurgenson E D, Navr\'atil P and Furnstahl R J 2009 arXiv:nucl-th/09125.1873
\bibitem{Gazit09} Gazit D, Quaglioni S and Navr\'atil P 2009 {\it Phys. Rev. Lett.}
{\bf 103} 102502
\bibitem{Covello05} Covello A, Coraggio L, Gargano A and Itaco N 2005 {\it J. Phys. Conf. Ser.} {\bf 20} 137
\bibitem{Barrett70} Barrett B R and Kirson M W 1970 {\it Nucl. Phys.} A  {\bf 148} 145
\bibitem{Sommermann81} Sommermann H M, M\"uther H, Tam K C, Kuo T T S and Faessler A 1981 {\it Phys. Rev. } C {\bf 23} 1765
\bibitem{Schucan72} Schucan T H, Weidenm\"uller H A 1972 {\it Ann. Phys. N.Y.} {\bf 73} 108\bibitem{Shurpin83} Shurpin J, Kuo T T S and Strottman D 1983 {\it Nucl. Phys. } A {\bf 408} 310
\bibitem{Hjorth95} Hjorth-Jensen M, Kuo T T S and Osnes E 1995 {\it Phys. Rep.} {\bf 261} 125
\bibitem{Holt05} Holt J D, Holt J W,  Kuo T T S, Brown G E and Bogner S K 2005 {\it Phys. Rev. } C {\bf 72} 041304(R)
\bibitem{Zuker03} Zuker A P 2003 {\it Phys. Rev. Lett.} {\bf 90} 042502
\bibitem{Ellis05} Ellis P J, Engeland T, Hjorth-Jensen M, Kartamyshev M P and Osnes E
2005 {\it Phys. Rev. } C {\bf 71} 034301
\bibitem{Coraggio09d} Coraggio L, Covello A, Gargano A and Itaco N 2009 {\it Phys. Rev. } C
{\bf 80} 044311
\bibitem{Coraggio09e} Coraggio L, Covello A, Gargano A, Itaco N and Kuo T T S 2009 {\it Phys. Rev. } C {\bf 80} 044320
\bibitem{Coraggio09b} Coraggio L, Covello A, Gargano A and Itaco N 2009 {\it Phys. Rev.} C
{\bf 80} 021305(R)
\bibitem{Coraggio09c} Coraggio L, Covello A, Gargano A and Itaco N 2009 {\it Phys. Rev. } C
{\bf 80} 061303(R)
\bibitem{Covello07} Covello A, Coraggio L, Gargano A and Itaco N 2007 {\it Prog. Part. Nucl. Phys.} {\bf 59} 401
\bibitem{Simpson07} Simpson G S {\it et al.} 2007  {\it Phys. Rev. } C {\bf 76} 041303(R)
\bibitem{Liu09} Liu S H {\it et al.} 2009 {\it Phys. Rev. } C {\bf 79} 067303
\bibitem{Covello97} Covello A, Andreozzi F, Coraggio L, Gargano A, Kuo T T S and Porrino A
1997 {\it Prog. Part. Nucl. Phys.} {\bf 38} 165
\bibitem{Holt97} Holt A, Engeland T, Osnes E, Hjorth-Jensen M and Suhonen J 1997 
{Nucl. Phys.} A {\bf 628} 107
\bibitem{Engeland98} {Engeland T, Holt A, Osnes E and Hjorth-Jensen M 1999} {\it Proc. 6th Int. Spring Seminar on Nuclear Physics (S. Agata sui due Golfi, 1998)} ed A. Covello (Singapore: World Scientific) p 117
\bibitem{Coraggio07} Coraggio L, Covello A, Gargano A, Itaco N, Entem D R, Kuo T T S and Machleidt R 2007 {\it Phys. Rev.} C {\bf 75} 024311
\bibitem{Otsuka09} Otsuka T, Suzuki T, Holt J D, Schwenk A and Akaishi Y 2009 arXiv:nucl-th/0908.2607
\bibitem{Genevey03} Genevey J {\it et al.} 2003 {\it Phys. Rev. } C {\bf 67} 054312
\bibitem{Scherillo04} Scherillo A {\it et al.} 2004 {\it Phys. Rev. } C {\bf 70} 054318
\bibitem{Tsukiyama09} Tsukiyama K, Hjorth-Jensen M and Hagen G 2009 {\it Phys. Rev.} C
{\bf 80} 051301



\end{thebibliography}
\end{document}